\documentclass{JHEP3}
\usepackage[T1]{fontenc}
\usepackage{amsfonts}
\usepackage{amssymb}
\usepackage{amsmath}
\usepackage[dvips]{graphicx,epsfig}
\usepackage{verbatim}

\def\braket#1{\mathinner{\langle{#1}\rangle}}

{\catcode`\|=\active 
  \gdef\Braket#1{\left<\mathcode`\|"8000\let|\BraVert {#1}\right>}}
\def\BraVert{\egroup\,\mid@vertical\,\bgroup}
{\catcode`\|=\active
  \gdef\set#1{\mathinner{\lbrace\,{\mathcode`\|"8000\let|\midvert #1}\,\rbrace}}
  \gdef\Set#1{\left\{\:{\mathcode`\|"8000\let|\SetVert #1}\:\right\}}}
\def\midvert{\egroup\mid\bgroup}
\def\SetVert{\egroup\;\mid@vertical\;\bgroup}
\newcommand{\R}{\mathbb{R}}

\newcommand{\tr}{\mathrm{Tr}}
\newcommand{\comm}[2]{\left[ #1 , #2 \right]}

\newcommand{\Jb}{\bar{J}}

\title{\bf Quantum Fields with Noncommutative Target Spaces}

\author{A. P. Balachandran$^a$, A. R. Queiroz$^{b}$, A. M. Marques$^{c}$ and P. Teotonio-Sobrinho$^{c}$ \\ $^a$Department of Physics, Syracuse University,
Syracuse, NY, 13244-1130, USA. \\ $^b$Centro Internacional de F\'{\i}sica da Mat\'{e}ria Condensada,
 Universidade de Bras\'{\i}lia,\\ C.P. 04667, Bras\'{\i}lia, DF, Brazil.  \\ $^c$Instituto de F\'{\i}sica, Universidade de S\~{a}o Paulo, C.P. 66318, 05315-970,\\ S\~{a}o Paulo, SP, Brazil.  \\ E-mail: \email{bal@physics.syr.edu}, \email{amilcar@fma.if.usp.br}, \email{amarques@fma.if.usp.br}, \email{teotonio@fma.if.usp.br}}

\preprint{SU-4252-851}
\abstract{Quantum field theories (QFT's) on noncommutative spacetimes are currently under intensive study. Usually such theories have world sheet noncommutativity. In the present work, instead, we study QFT's with commutative world sheet and noncommutative target space. Such noncommutativity can be interpreted in terms of twisted statistics and is related to earlier work of Oeckl \cite{Oeckl:1999zu}, and others \cite{Balachandran:2005eb,Balachandran:2005pn,Balachandran:2006ib,Balachandran:2006pi,Pinzul:2005gx,Fiore:2007vg,Sasai:2007me}. The twisted spectra of their free Hamiltonians has been found earlier by Carmona et al \cite{Carmona:2002iv,Carmona:2003kh}. We review their derivation and then compute the partition function of one such typical theory. It leads to a deformed black body spectrum, which is analysed in detail. The difference between the usual and the deformed black body spectrum appears in the region of high frequencies. Therefore we expect that the deformed black body radiation may potentially be used to compute a GZK cut-off which will depend on the noncommutative parameter $\theta$.}

\keywords{Noncommutative geometry, Conformal Field Theory, Noncommutative Target Space, GZK Cut-off}

\begin{document}
%\maketitle

\section{Introduction}

Spacetime noncommutativity is suggested by considerations based on quantum gravity \cite{Doplicher:1994tu} and developments in string theory \cite{Frohlich:1993es, Landi:1998ii, Seiberg:1999vs,Alekseev:1999bs,Douglas:2001ba}. Quantum field theories (qft's) on such spacetimes have been developed first Doplicher et al. \cite{Bahns:2002vm}, and Oeckl \cite{Oeckl:1999zu}. Studies of spacetime symmetries of such models \cite{Aschieri:2005zs, Chaichian:2004za} have also led to some insight into their properties, such as Pauli principle violation \cite{Balachandran:2005eb} and the absence of UV-IR mixing \cite{Balachandran:2005pn}. The earlier work of Oeckl \cite{Oeckl:2000eg} using the technology of Hopf algebras have much bearing on such topics and anticipated crucial results. But it is only recently that general attention has focused on the ideas underlying these results. 

In order to fully explore the consequences of these results, phenomenological models based on such noncommutative framework are useful. In this regard it would be highly desirable to measure at least a definitive signature of the fact that physics at a scale close to the Planck scale takes place in a noncommutative geometry. One usually expects that signatures of noncommutativity will appear in experiments involving cosmic microwave background (CMB), ultra-high-energy cosmic rays, or other high-energy sources such as those of neutrinos. It is well-known that the CMB radiation shows a black body spectrum with great accuracy, at least for low to medium frequencies. Still we may conjecture that as the measurements become more accurate, deviations from the black body spectrum may be found, particularly in high frequency regions, where data are still not so accurate. Furthermore, the distribution of photons play a key role on the derivation of the GZK cut-off \cite{Geddes:1995sd}. Upto now estimates of the GZK cut-off are obtained by considering a black body distribution. Deviations from this cut-off may thus provide us signals of noncommutativity.

These facts convince us that any deviation from the usual free massless boson theory may have some influence on the modeling of these experimental facts. One possible source of such deviation is the deformation of the target space of a free massless boson theory. For instance, we may consider a theory with a noncommutative plane as target space. Then a natural question one may raise is how the radiation spectrum is modified by this deformation of the target-space of the free massless boson theory.

We present here a free massless boson theory with commutative base-space and noncommutative target-space. The first results with bearing on such models can be found in the work of Oeckl \cite{Oeckl:1999zu}, and can be interpreted in terms of twisted statistics. The idea of deformation of the target space has also been extensively developed in the work\footnote{ We thank Prof. J. Gamboa and Prof. J. Cortes for calling our attention to their interesting earlier work.} of Carmona et al \cite{Carmona:2002iv,Carmona:2003kh}. These works use the earlier work on quantum mechanics with noncommutative target spaces by Nair and Polychronakos \cite{Nair:2000ii}. Their physical consequences have also been explored by the same group in \cite{Das:2004vc,Gamboa:2005pd,Falomir:2005it,Gamboa:2005bf,Arias:2006fu,Arias:2007xt}. They call the fields with noncommutative target spaces as noncommutative fields. Subsequently Balachandran et al \cite{Balachandran:2005eb,Balachandran:2005pn,Balachandran:2006ib,Balachandran:2006pi,Balachandran:2006hr} developed and extended the ideas of Oeckl, with emphasis on the link between twisted Poincaré invariance and twisted statistics. Note that it can be argued that spacetime and target space noncommutativity are linked to each other by symmetry principles.

Our treatment of deformed conformal symmetry in section 2.3 is new. Also unlike Carmona et al, we consider nonconstant deformation parameters and discuss the deformed black body radiation spectrum in detail. This study can acquire significance from the point of view of of phenomenology. 

Sections 2 and 3 contain many results already obtained by \cite{Carmona:2002iv,Carmona:2003kh}, although perhaps our derivation of their results is slighty different. They are included here for completeness.

\begin{comment}
There are some differences between our approach and the approach of Carmona et al worthy mentioning. One difference is in the algebra of fields and and their conjugated momenta. In their work, they deform the commutation relation between canonical momenta also, besides the deformation of the commutation relation between the fields. In our work, we keep undeformed the commutation relation between momenta, as can be seen ahead in (\ref{eq:COMMPHIPI}). Furthermore, we consider a regularisation of the delta function, or equivalently a non constant $\theta$, as can be seen ahead in (\ref{eq:THETAS}). This regularisation allows us to further analyse the new energy density radiation and discuss its property compared with the usual black body radiation.
\end{comment}

The present work is organised as follows. In section 2, we construct a free  massless boson theory with noncommutative $\R^2$, hereafter referred to as the deformed theory, for the case where the base space is a $(1+1)$-d space given by the cylinder $S^1\times \R$. After constructing the Hamiltonian formulation of this theory and quantizing it, we calculate $2$-point correlation functions and also see how the conformal generators are modified by the introduction of noncommutativity. The results obtained in this section are, then, generalised to a more realistic scenario in section 3. In this section, we show the same theory, but now with the $(3+1)$-d base space given by $S^1\times S^1\times S^1\times \R$. Then, using this $(3+1)$-d theory, we study the radiation law for the deformed theory and compare it with the usual commutative theory, i.e., the black body radiation. Concluding remarks are section 5.

\section{Deformed $(1+1)$-d Theory}

In this section we construct a deformed version of a $(1+1)$-d theory. For that we start with a target space $\R^2$ as
a group manifold. Then we use a twisting procedure on the group $\R^2$, so that it becomes the
noncommutative plane. 

The phase space $T^*\R^2$ of a single particle in $\R^2$ has a natural group structure. It is the semi-direct product of $\R^2$ with $\R^2$. The generators of its Lie algebra can be taken to be coordinates $x^{a},a=1,2$, and momenta $p_{a}$. The algebra generated by $x^a$ and $p_a$ are given by
\begin{eqnarray}
\comm{x^{a}}{x^{b}}&=&0 \\
\comm{p_{a}}{p_{b}}&=&0 \\
\comm{x^{a}}{p_{b}}&=&i\delta^a_b\,.
\end{eqnarray}
Functions on $\R^2$ are thus commutative and $p_a$ acts as derivative on these functions.

We now twist (or deform) the generators of the above algebra into $\hat{x}^{a},{\hat{p}}_{b}$ and thus obtain a new algebra.
The algebra of derivatives are not deformed, but the function algebra is.
The twisted algebra reads \cite{Nair:2000ct,Nair:2000ii}
\begin{eqnarray}
\comm{\widehat{x}^{a}}{\widehat{x}^{b}}&=&i\epsilon^{ab}\theta\equiv i\theta^{ab} \label{eq:DefCR1}\\
\comm{\widehat{p}_{a}}{\widehat{p}_{b}}&=&0 \label{eq:DefCR2}\\
\comm{\widehat{x}^{a}}{\widehat{p}_{b}}&=&i\delta^{a}_b. \label{eq:DefCR3}
\end{eqnarray}
The so-called non-commutativity parameter $\theta^{ab}=-\theta^{ba}$ is as usual taken
as a constant for the present.

We can relate the hatted generators (noncommutative ones) to the usual generators by the deformation map\footnote{The analogue of the map $x^a\to\hat{x}^a$ first appeared in the works of Grosse \cite{1979PhLB86267G}, Zamolodchikov and Zamolodchikov \cite{Zamolodchikov:1978xm} and Faddeev \cite{Faddeev:1980zy}. It is sometimes called a dressing transformation.}
\begin{eqnarray}
\label{eq:Bal-map}
\widehat{x}^{a}&=&x^{a}-\frac{1}{2}\theta^{ab}p_{b}, \nonumber \\
\widehat{p}^{a}&=&p^{a}.
\end{eqnarray}
In what follows, we will generalize the above deformation map to the case of a scalar field theory.

We start with a free real massless bosonic field. Its base space is a cylinder with
circumference $R$ and its target space is $\R^2$, i.e.,
\[
\begin{array}{cccl}
\varphi:&S^1 \times \R &\longrightarrow& \R^2 \\
        &(x,t)         &\longmapsto    & \varphi(x,t) 
\end{array}
\]
The field components will be denoted
$\varphi^i$ where $i = 1,2$. The compactification of the space
coordinate makes each field component periodic in the variable $x$, \emph{i.e.}, 
\[ \varphi^i(x+R,t) \equiv \varphi^i(x,t)\,. \]

The field components $\varphi^i(x,t)$ may be written as a Fourier expansion
\begin{equation}
  \label{eq:FEXPANSION}
  \varphi^i(x,t) = \sum_n \;e^{\frac{2\pi i}{R}n x} \; \varphi^i_n(t)\;,
\end{equation}
where the Fourier components are
\begin{equation}
  \label{eq:FTRANSFORM}
  \varphi^i_n(t) = \frac{1}{R} \int dx \; e^{-\frac{2\pi i}{R}n x} \; \varphi^i(x,t)\;=\varphi_{-n}^i(t)^*.
\end{equation}

The Lagrangian for the free field is
\begin{equation}
  \label{eq:FREEFIELDLAGRANGIAN}
L = \frac{g}{2} \int dx \big[ \big(\partial_t \varphi^i \big)^2 - \big(
  \partial_x \varphi^i \big)^2 \big] 
\end{equation}
where there is an implicit sum over the target space coordinate index $i$.

In terms of Fourier modes, the above Lagrangian is 
\begin{equation}\label{eq:LFMODES}
  L =  \frac{gR}{2} \sum_{i,n} \bigg\{\dot{\varphi}^i_n \dot{\varphi}^i_{-n} -
  \bigg( \frac{2\pi n}{R}\bigg)^2 \varphi^i_n \varphi^i_{-n} \bigg\}\;.
\end{equation}

It is a standard exercise to check that the theory defined by the Lagrangian
(\ref{eq:FREEFIELDLAGRANGIAN}) possesses conformal invariance. This is
verified by showing that the energy-momentum tensor is traceless.

We can define the momenta conjugate to $\varphi^i_n=\varphi_{-n}^{i~*}$ as
\begin{equation}
  \label{eq:MOMENTA}
  \pi^i_n = \frac{\partial L}{\partial \dot{\varphi}^i_n} = gR\,\dot{\varphi}^i_{-n} = \pi_{-n}^{i~*}.
\end{equation}
We then have the following commutation relations:
\begin{equation}
  \label{eq:COMMPHIPI}
\comm{\varphi_m^i}{\varphi_n^j} = \comm{\pi_m^i}{\pi_n^j} = 0 \hspace{1cm}  \comm{\varphi_m^i}{\pi_n^j} = i \delta_{mn}\delta^{ij}\;.
\end{equation}

The Hamiltonian for (\ref{eq:LFMODES}) is 
\begin{equation}\label{eq:HAMZERO}
   H = \sum_i \frac{(\pi_0^i)^2}{2gR} + \frac{1}{2gR} \sum_{i, n \neq 0} \big\{ \pi^i_n \pi^i_{-n} + (2 \pi |n| g )^2 \varphi^i_n
  \varphi^i_{-n} \big\}\;.
\end{equation}
For $n\neq 0$, it is equivalent to an infinite set of decoupled harmonic
oscillators with frequencies
\begin{equation}
  \label{eq:OMEGAN}
  \omega_n = \frac{2\pi |n|}{R}\; .
\end{equation}
The first term in the l.h.s. of equation
(\ref{eq:HAMZERO}) is known as the \emph{zero mode}. We will mostly ignore it in the rest of these notes. It will not affect our conclusions.
The hermiticity of $\varphi^i$ implies that
\begin{equation}
  \label{eq:PHIDAGGER}
  \varphi_n^{i\;\dagger} (t) = \frac{1}{R} \int dx e^{\frac{2\pi i n x}{R}}
  \varphi^i(x,t) = \varphi^i_{-n}(t)
\end{equation}
and 
\begin{equation}
  \label{eq:PIDAGGER}
  \pi_n^{i\;\dagger} = \pi^i_{-n}\;.
\end{equation}

We now generalize the deformation map (\ref{eq:Bal-map}) to the above scalar field theory. We take
ordinary field theory with target $\R^2$ described by canonical
pairs $\varphi^{a},\pi_{a}$, $a=1,2$, and then use the map (\ref{eq:Bal-map}) so that
\begin{eqnarray}
\hat{\varphi}^{a}(x,t)&=&\varphi^{a}(x,t)-
\frac{1}{2}\theta^{ab} \pi_{b}(x,t), \label{eq:DEFFIELD} \\
\hat{\pi}_{a}(x,t)&=&\pi_{a}(x,t)\;. \label{eq:DEFMOM}
\end{eqnarray}
In the case where
$\theta^{ab}=\theta \epsilon^{ab}$, with $\theta$ chosen as a constant, the corresponding equal time commutation relations are
\begin{eqnarray}
\comm{\hat{\varphi}^{a}(x,t)}{\hat{\varphi}^{b}(y,t)}&=&i
\epsilon^{ab}\theta\delta(x-y), \label{eq:NCCR1}\\
\comm{\hat{\pi}_{a}(x,t)}{\hat{\pi}_{b}(y,t)}&=&0, \label{eq:NCCR2}\\
\comm{\hat{\varphi}^{a}(x,t)}{\hat{\pi}_{b}(y,t)}&=&i\,\delta^{a}_b\delta(x-y).\label{eq:NCCR3}
\end{eqnarray}
These are the relations of \cite{Carmona:2002iv,Carmona:2003kh} in the limit of their $B$ becoming zero.

In what follows, we are going to redefine $\theta\delta(x-y)$ such that
\begin{equation}\label{eq:THETAS}
\theta(\sigma)=\frac{\theta}{\sqrt{2\pi}\sigma}e^{-\frac{(x-y)^2}{2\sigma^2}}
\end{equation}
which is the previous constant $\theta$ times a function of a new
parameter $\sigma$. This new parameter $\sigma$ introduces a new distance scale
in the  equal time commutation relations for the fields:
\begin{equation}\label{eq:NCCRSIGMA}
\comm{\hat{\varphi}^{a}(x,t)}{\hat{\varphi}^{b}(y,t)}=\frac{i
\epsilon^{ab}\theta}{\sqrt{2\pi}\sigma}e^{-\frac{(x-y)^2}{2\sigma^2}}\;.\\ 
\end{equation}

It is important to note that these new commutation relations reduces to
those in (\ref{eq:NCCR1}) in the limit $\sigma \rightarrow 0$.
We note further the fact that while the new $\theta(\sigma)$ is
dimensionless,  because $\theta(\sigma)$ substitutes the previous $\theta\delta(x-y)$,
the $\theta$ in $\theta(\sigma)$ still has the dimension of $(\textrm{length})^{-2}$.

In more than two dimensions, it is not easy to introduce spacetime dependence in the noncommutative parameter (called also $\theta^{\mu \nu}$) of the Moyal plane \cite{Bal-book-2}, because of constraints from Jacobi identity. But here we twist the target of fields, so that this problem does not arise.

The above deformed algebra describes a field with noncommutative plane as target. The degree
of non-commutativity is regulated by a new parameter $\sigma$.

We can describe the deformed algebra in terms of the Fourier modes
\begin{equation}
\label{FModes-1}
\hat{\varphi}^{a}=\sum_{n}e^{\frac{2\pi i}{R}nx}\hat{\varphi}_{n}^{a}.
\end{equation}
Then the ``dressing transformation''
\begin{eqnarray}
\hat{\varphi}_{n}^{a}&=&\varphi_{n}^{a}-\frac{1}{2R}\epsilon^{ab}\theta(n)\pi_{-n}^{b}, \label{eq:fmodes-a}\\
\hat{\pi}_n^{a}&=&\pi_n^{a} \label{eq:fmodes-b},
\end{eqnarray}
where $R$ is the circumference of the base space cylinder $S^1\times \R$ and
\begin{equation}
  \label{eq:THETAN}
  \theta(n)=\theta\; e^{-\frac{2\pi^2\sigma^2 n^2}{R^2}}\,.
\end{equation}
reproduces the Fourier components of (\ref{eq:NCCRSIGMA}), (\ref{eq:NCCR2}) and (\ref{eq:NCCR3}):
\begin{eqnarray}
 \comm{\hat{\varphi}_n^{a}}{\hat{\varphi}_m^{b}}&=&\frac{i}{R}\epsilon^{ab}\theta(n)\delta_{n+m,0}, \label{eq:FNCCR1}\\   
\comm{\hat{\pi}_n^{a}}{\hat{\pi}_m^{b}}&=&0, \label{eq:FNCCR2}\\
\comm{\hat{\varphi}_n^{a}}{\hat{\pi}_m^{b}}&=&i\,\delta^{ab} \delta_{mn}. \label{eq:FNCCR3}
\end{eqnarray}
(Notice that $[\theta(n)]=\textrm{dimension of }\theta(n)=\textrm{length}$.)
% Target space is, again, $\R^2$.
% Deformed fields and conjugated momenta:
% \begin{eqnarray}
% %%\TAG{eq:DEFPHI}
%   \hat{\varphi}^a_n &=& \varphi^a_n + \theta^{ab} \pi_{-n}^b \label{eq:DEFPHI}  \\
% %%\TAG{eq:DEFPI}
%   \hat{\pi}^a_n &=& \pi^a_n \label{eq:DEFPI}
% \end{eqnarray}
% where $\theta_{ab} = \theta \, \epsilon_{ab}$. Expanding (\ref{eq:DEFPHI})
% and (\ref{eq:DEFPI}) using (\ref{eq:FEXPANSION2}) and (\ref{eq:FTRANSFORM2})
% we get
% \begin{equation}
%   \label{eq:DEFFIELDS}
%   \hat{\varphi}^a(x,t) = \varphi^a(x,t) + \theta_{ab}\,\pi^b(x,t)
%   \hspace{5mm}\mathrm{and}\hspace{5mm}
%   \hat{\pi}^a(x,t) = \pi^a(x,t)
% \end{equation}
% with
% \begin{eqnarray}
%   \label{eq:DEFCR}
%   \comm{\hat{\varphi}^a(x,t)}{\hat{\varphi}^b(y,t)} &=& i\,\theta^{ab}\,\delta(x-y) \\
%   \comm{\hat{\varphi}^a(x,t)}{\pi^b(y,t)} &=& i \,\delta^{ab} \,
%   \delta(x-y) \\
%   \comm{\pi^a(x,t)}{\pi^b(y,t)} &=& 0  
% \end{eqnarray}

We can write the Hamiltonian in terms of the deformed fields and momenta,
using equations (\ref{eq:DEFFIELD}), (\ref{eq:fmodes-a}) and (\ref{eq:fmodes-b}). Then, the Hamiltonian, without the zero mode term, reads
\begin{eqnarray}
  H &=& \frac{1}{2gR}\sum_{i,n} \left\{ \hat{\pi}_n^i\,\hat{\pi}_{-n}^i + (2\pi |n| g)^2
  \, \hat{\varphi}^i_n \, \hat{\varphi}^i_{-n} \right\} \nonumber \\  
% &=& \frac{1}{2gR}\sum_{(i,n)} \; \pi_n^i\,\pi_{-n}^i + (2\pi n g)^2
%  (\varphi^i_n - \epsilon^{ij}\theta(n)R^{-1}\pi_{-n}^j)(\varphi^i_{-n} -
%  \epsilon^{ij}\theta(n)R^{-1}\pi_{n}^j)  \nonumber \\
&=& \frac{1}{2gR}  \sum_{i,j,n}  \left\{ \left[1+\left(\frac{\pi g |n|
      \theta(n)}{R}\right)^2\right]\,\pi_n^i \, \pi_{-n}^i  \right. +  \nonumber \\
& &  \left. \phantom{\frac{1}{2gR}  \sum_{(i,k,n)} \;} 
+ (2\pi |n| g)^2 \, \varphi_n^i \, \varphi_{-n}^i -\frac{(2\pi |n| g)^2}{R}
\epsilon_{ij}\theta(n)\varphi_n^i \pi_n^j \right\} \;.   \label{eq:HAMPHIPI}
\end{eqnarray}

\subsection{The Schwinger Representation and the Construction of the Fock Space} 

We can construct a Fock  space, since the Hamiltonian
(\ref{eq:HAMPHIPI}) can  be 
diagonalised using the Schwinger representation of $SU(2)$. 
It is remarkable that the last term of the Hamiltonian 
in equation  (\ref{eq:HAMPHIPI}) is proportional to  the $z$-component of the angular momentum $L_z$. This fact motivates the
use of the Schwinger representation.

First, let us rewrite the full Hamiltonian (\ref{eq:HAMPHIPI}) as
\begin{equation}
  \label{eq:HAMPHIPI2}
  H = \sum_i \frac{(\pi_0^i)^2}{2gR}+\sum_{i,j,n\neq 0} \left( \; \frac{\Omega^2_n}{2gR}\;\pi_n^i \,\pi_{-n}^i\,
 + \frac{gR}{2} \; \omega_n^2 \, \varphi_n^i \,
\varphi_{-n}^i - \frac{g}{2}\theta(n)\,\omega_n^2 \,\epsilon_{ij}\varphi_n^i \pi_n^j \;\right),  
\end{equation}
where
\begin{equation}
  \Omega^2_n = 1+\left(\frac{\pi g |n|\theta(n)}{R}\right)^2\hspace{10mm}\mathrm{and}\hspace{10mm}\omega_n = \frac{2
    \pi |n|}{R}\;.
\end{equation}
It is important to note that in the limit $\theta\rightarrow 0$, we have
$\theta(n)\rightarrow 0$ and $\Omega^2_n \rightarrow 1$. This limit takes the
Hamiltonian (\ref{eq:HAMPHIPI2}) to the usual Hamiltonian for the free massless bosons upon a commutative target space.

Separating the zero mode contribution $H_0$, we have
\begin{equation}
  H = H_0+\sum_{i,j,n\neq 0} \left( \frac{\Omega^2_n}{2gR}\, \pi_n^i \,\pi_{-n}^i
 + \frac{gR}{2} \; \omega_n^2 \, \varphi_n^i \,
\varphi_{-n}^i -\frac{g}{2}\theta(n)\,\omega^2 \,\epsilon_{ik}\,\varphi_n^i \pi_n^k \right).
\end{equation}
Therefore, the Hamiltonian may be written as
\begin{equation}
  H = H_0 +  \sum_{n\neq 0} \; \big( {H}_n - \frac{g}{2}\,\theta(n) \omega_n^2 \,J_n^z\; \big),
\end{equation}
where
\begin{equation}
  {H_0} =  \sum_i \frac{{\pi_0^i}^2}{2gR}\;,
\end{equation}
\begin{equation}
H_n = \sum_i \left(\frac{\Omega^2_n}{2gR}\,\pi_n^i \,\pi_{-n}^i
 + \frac{gR}{2}\,\omega_n^2 \, \varphi_n^i \,
\varphi_{-n}^i\right)
\end{equation}
and
\begin{equation}
  \label{eq:Jz}
  J_n^z = \sum_{i,j} \epsilon_{ij}\, \varphi_n^i \, \pi_n^j\;.
\end{equation}

${H}_n$ alone is the Hamiltonian of a harmonic oscillator with frequency
$\Omega_n \omega_n$, which can be easily diagonalised if we note that it can be
rewritten as
\begin{equation}
  H_n = \sum_i \left(\frac{1}{2M}\,\pi_n^i \,\pi_{-n}^i + \frac{1}{2}\,M
  \bar{\omega}_n^2 \, \varphi_n^i \,\varphi_{-n}^i\right).
\end{equation}
This is the traditional form of the harmonic oscillator Hamiltonian, where
\begin{equation}
  M = \frac{gR}{\Omega^2_n}
  \hspace{1cm}\mathrm{and}\hspace{1cm}\bar{\omega}_n= \Omega_n\,\omega_n.
\end{equation}
We may now define annihilation and creation
operators. For $n \neq 0$:
\begin{equation}
  a_n^i = \sqrt{\frac{\Delta_n}{2}} \bigg( \varphi_n^i + i \frac{\pi_{-n}^i}{\Delta_n
  }\bigg) \hspace{10mm}  {a_{n}^i}^\dagger = \sqrt{\frac{\Delta_n}{2}} \bigg(
  \varphi_{-n}^i - i \frac{\pi_{n}^i}{\Delta_n}\bigg)
\end{equation}
where
\begin{equation}
  \Delta_n = M\bar{\omega}_n= \frac{g R \omega_n}{\Omega_n} =  \frac{2 \pi \,|n| \, g}{\Omega_n}\;,
\end{equation}
and, as usual,
\begin{equation}
    \comm{a^i_m}{a^j_n} = \comm{{a^i_m}^\dagger}{{a^j_n}^\dagger} = 0 \hspace{1cm}
  \comm{a^i_m}{{a^j_n}^\dagger} = \delta_{mn}\delta^{ij}\;
\end{equation}
It is now easy to check that
\begin{equation}\label{eq:141}
  \varphi_n^i = \frac{1}{\sqrt{2\Delta_n}} (a_n^i + {a_{-n}^i}^\dagger)  \hspace{10mm}
  \pi_n^i = - i \sqrt{\frac{\Delta_n}{2}} ({a}_{-n}^i - {a_n^i}^\dagger)
\end{equation}
and that if we make $\theta \rightarrow 0$, we have $\Omega_n \rightarrow 1$
and $\Delta_n \rightarrow 2\pi|n|g$ and in this limit $a_n^i $ and
${a_{n}^i}^\dagger$  reduce to their usual (commutative space-time)
version.

Now, in terms of these annihilation and creation
operators the Hamiltonian becomes
\begin{equation}
  {H}_n = \sum_i  \frac{\omega_n \,\Omega_n}{2} \; ( {a_n^i}\,
  {a_n^i}^\dagger + {a_{-n}^i}^\dagger \, {a_{-n}^i})
\end{equation}
which, using some algebra, can be written as (the $:\ :$ symbols stand for the usual normal ordering of operators)
\begin{eqnarray}
  \sum_{n\neq 0} H_n &=&  \sum_{i,n \neq 0}  \frac{\omega_n \, \Omega_n}{2} :( {a_n^i}\,
  {a_n^i}^\dagger + {a_{-n}^i}^\dagger \, {a_{-n}^i}): \nonumber \\ &&
  \nonumber \\
&=&  \sum_{i,n \neq 0}
  \frac{\omega_n \, \Omega_n}{2} \; ({a_n^i}^\dagger \,{a_n^i} +
  {a_{-n}^i}^\dagger \, {a_{-n}^i}) \nonumber \\ && \nonumber \\
&=& \sum_{i,n \neq 0} 
  \omega_n \, \Omega_n \; {a_n^i}^\dagger \,{a_n^i}.
% \nonumber \\ && \nonumber \\
% &=& \sum_{i,n > 0}
%   \frac{\omega_n}{\sqrt{\Onsq}} \; N_n^i
\end{eqnarray}
% \[ \ket{n^i_m} = \frac{({a^\dagger}^i_m)^n}{\sqrt{n!}}\; \ket{\mathbf{0}}
% \]

% \[  N_n^i \ket{n^i_m} = n^i_m \ket{n^i_m} \]

% The most general state:

% \[ \ket{{n_1}_{m_1}^{i_1} , \ldots ,{n_i}_{m_i}^{i_i}, \ldots} =
% \bigotimes_{m\neq 0  \atop\ i = 1,2} \ket{n^i_m} \] 
Accordingly, for the angular momentum term we get
\begin{equation}
\sum_{n\neq 0}  J_n^z = - \sum_{n\neq0} i\;\epsilon_{ij}\;{a_{n}^i}^\dagger{a_{n}^j}\,, 
%= i\,\theta\,(a_1 a_2^\dagger - a_2 a_1^\dagger)  
\end{equation}
so that we can finally write the complete Hamiltonian as
\begin{equation}
  H = \sum_i \frac{(\pi^i_0)^2}{2gR} + \sum_{n\neq0}
  \omega_n \, \Omega_n\;{a_n^i}^\dagger a_n^i - i \frac{g}{2}\,\theta(n)\,\omega_n^2
  \epsilon_{ij}\,{a_{n}^i}^\dagger{a_{n}^j}\,.
\end{equation}

The appearance of the angular momentum term may be seen as a signature of the well-known relation between noncommutative plane and a plane embedded in a magnetic field perpendicular to it.

Now we define new annihilation and creation operators 
\begin{equation}\label{eq:117}
A^1_n = \frac{1}{\sqrt{2}} (a_n^1 - i a^2_n) \hspace{2cm}  A^2_n =
\frac{1}{\sqrt{2}} (a_n^1 + i a^2_n) 
\end{equation}
such that
\[ \comm{A^i_n}{{A^j_m}^\dagger} = \delta_{nm}\delta^{ij}\;. \]

Using equations (\ref{eq:141}) and (\ref{eq:117}), we can check that the field
components can be written as
\begin{eqnarray}
\varphi_n^1&=&\frac{1}{2\sqrt{\Delta_n}} (A_n^1 + A_{-n}^{1\dagger} + A_n^2
+ A_{-n}^{2\dagger}) \label{PHI1}\\
 \varphi_n^2&=&\frac{i}{2\sqrt{\Delta_n}} (A_n^1 - A_{-n}^{1\dagger} - A_n^2
+ A_{-n}^{2\dagger})\;. \label{PHI2} 
\end{eqnarray}

In terms of these $A$'s and $A^\dagger$'s the Hamiltonian reads 
 \begin{eqnarray}
  H &=&H_0+ 
%\frac{(\Pi_0)^2 + (\tilde{\Pi}_0)^2}{4gR} + 
\sum_{n \neq 0} \omega_n 
\bigg\{ 
\left( \Omega_n  + \frac{\pi|n| g \theta(n)}{R} \right) {A_n^1}^\dagger A_n^1 +
\left( \Omega_n  - \frac{\pi|n| g \theta(n)}{R} \right) {A_n^2}^\dagger A_n^2
\bigg\} \nonumber \\
 &=& 
%\frac{(\Pi_0)^2 + (\tilde{\Pi}_0)^2}{4gR} + 
\sum_{n \neq 0} \omega_n 
\bigg\{ 
\Lambda^1_n \, {A_n^1}^\dagger A_n^1 +
\Lambda^2_n \, {A_n^2}^\dagger A_n^2
\bigg\}
\end{eqnarray}
where
\begin{eqnarray}
  \Lambda_n^1 &=&  \Omega_n  + \frac{\pi|n| g \theta(n)}{R}, \\
  \Lambda_n^2 &=&  \Omega_n  - \frac{\pi|n| g \theta(n)}{R}\;.
\end{eqnarray}

Notice that an important effect of introducing noncommutativity is the
splitting of the energy levels of each individual mode that
constitutes the whole system. One should recall the resemblance of this effect to the well-known Zeeman effect in a quantum system in the presence of a magnetic field.

Time evolution of the  $A_n$'s and $A_n^\dagger$'s is given by
%\begin{equation}
%  \Phi(t) = \Phi^1_0 + \frac{\Pi_0}{gR}\,t\;, \hspace{15mm}
%  \tilde{\Phi}(t) = \tilde{\Phi}_0 + \frac{\tilde{\Pi}_0}{gR}\,t\;, \hspace{15mm}
%\end{equation}
\begin{equation}
  A_n^1(t) = A^1_n(0)\,e^{-i\omega_n\Lambda^1_n\,t}\;, \hspace{15mm}
  A_n^2(t) = A^2_n(0)\,e^{-i\omega_n\Lambda^2_n\,t}\;,
\end{equation}
\begin{equation}
  {A_m^1}^\dagger(t) = {A^1_m}^\dagger(0)\,e^{i\omega_m\Lambda^1_m\,t}\;, \hspace{15mm}
  {A_m^2}^\dagger(t) = {A^2_m}^\dagger(0)\,e^{i\omega_m\Lambda^2_m\,t}\;.
\end{equation}
which together with equation (\ref{PHI1}) and (\ref{PHI2}) gives, (omitting zero modes), 
\begin{eqnarray}
  \varphi^1(x,t) &=& \frac{1}{\sqrt{8\pi g}}
\sum_{n \neq 0} \sqrt{\frac{\Omega_n}{|n|}} \bigg\{ 
A^1_n          \,e^{ \frac{2\pi i}{R}(nx - |n| \Lambda^1_n t)} +
A^{1\dagger}_n \,e^{-\frac{2\pi i}{R}(nx - |n| \Lambda^1_n t)} +
\nonumber \\ && \nonumber \\
&& \phantom{\frac{1}{\sqrt{8\pi g}} \sum_{n \neq 0}
   \bigg\{}
+ A^2_n\,e^{\frac{2\pi i}{R}(nx - |n| \Lambda^2_nt)} +  
A^{2\dagger}_{n}\,e^{-\frac{2\pi i}{R}(nx - |n| \Lambda^2_nt)} \bigg\} \label{eq:PHI1D}
\end{eqnarray}
and
\begin{eqnarray}
  \varphi^2(x,t) &=& \frac{i}{\sqrt{8\pi g}} 
\sum_{n \neq 0} \sqrt{\frac{\Omega_n}{|n|}} \bigg\{ 
A^1_n            \,e^{\frac{2\pi i}{R}(nx - |n|\Lambda^1_n t)} -
A^{1\dagger}_{n}\,e^{-\frac{2\pi i}{R}(nx - |n|\Lambda^1_n  t)} -
\nonumber \\ && \nonumber \\
&& \phantom{\frac{1}{\sqrt{8\pi g}} \sum_{n \neq 0} \bigg\{}
- A^2_n         \,e^{\frac{2\pi i}{R}(nx - |n| \Lambda^2_nt)} +  
A^{2\dagger}_{n}\,e^{-\frac{2\pi i}{R}(nx - |n| \Lambda^2_nt)} 
\bigg\} \label{eq:PHI2D}
\end{eqnarray}

\subsection{Correlation Functions}

Given the new fields (\ref{eq:PHI1D}) and (\ref{eq:PHI2D})
we can evaluate the two-point correlation functions, e.g., for 
$\varphi^1$:
\begin{eqnarray}
  \label{eq:71}
  \braket{\varphi^1(x_1,t_1)\varphi^1(x_2, t_2)} &=& \frac{1}{8\pi g} \sum_{n \neq
  0} \frac{\Omega_n}{n} \;  e^{\frac{2\pi i n x}{R}(x_1 - x_2)} \times
\nonumber \\ & & \nonumber \\ && \phantom{\frac{1}{8\pi g}}
\times \left[ e^{-\frac{2\pi i}{R} |n| \Lambda^1_n (t_1 - t_2)} +  e^{-\frac{2\pi i}{R} |n| \Lambda^2_n (t_1 - t_2)}\right].
\end{eqnarray}
Taking the limit $\theta \rightarrow 0$ of this expression we get
\begin{equation}
  \label{eq:72}
  \braket{\varphi^1(x_1,t_1)\varphi^1(x_2, t_2)} = \frac{1}{4\pi g} \sum_{n \neq
  0} \frac{1}{n} \; e^{\frac{2\pi i n x}{R}(x_1 - x_2)} \; e^{-i \omega_n
  (t_1 -t_2)}\;,
\end{equation}
which is exactly what we would get for the two-point correlation function on the commutative plane.

The same calculation can be performed for $\varphi^2(x,t)$ with a similar
result. Furthermore:
\begin{equation}
  \label{eq:72-b}
  \braket{\varphi^1(x_1,t_1)\varphi^2(x_2, t_2)} = 0\;.
\end{equation}

\subsection{The Deformed Conformal Generators}

We study in the sequel how the conformal generators are deformed by the deformations introduced by the noncommutativity of the target space.
 
The deformed Hamiltonian is written in terms of the hatted operators as
\begin{equation}
  H = \frac{1}{2gR}\sum_i (\pi_0^i)^2+\frac{1}{2gR}\sum_{i,n} \left( \hat{\pi}_n^i\,\hat{\pi}_{-n}^i + (2\pi |n| g)^2
  \, \hat{\varphi}^i_n \, \hat{\varphi}^i_{-n} \right). 
\end{equation}
It can be diagonalised if we define the deformed creation and annihilation operators as
\begin{eqnarray}
  {\hat{a}^i_n} & = & \frac{1}{\sqrt{4\pi |n| g}}\big( 2\pi g |n| \hat{\varphi}^i_n +
  i \hat{\pi}_{-n}^i \big), \label{AH} \\
  {\hat{a}^{i\dagger}_n} &=& \frac{1}{\sqrt{4\pi |n| g}}\big( 2\pi g |n| \hat{\varphi}^i_{-n} -
  i \hat{\pi}_{n}^i \big). \label{AHD}
\end{eqnarray}
Using the commutation relations (\ref{eq:FNCCR1}), (\ref{eq:FNCCR2}) and (\ref{eq:FNCCR3}), we obtain the commutation relations
\begin{eqnarray}
  \comm{\hat{a}^i_m}{\hat{a}^j_n} =
  \comm{\hat{a}^{i\dagger}_m}{{\hat{a}^{j\dagger}_n}} &=&  \frac{i \pi g |m|}{R}
  \theta(n)\epsilon^{ij}\delta_{m+n}, \label{def-osc-1} \\  &&  \nonumber \\ \label{def-osc-2}
  \comm{\hat{a}^i_m}{{\hat{a}^{j\dagger}_n}} &=& \frac{ i \pi g |m| }{R} \theta(n)\epsilon^{ij}
  \delta_{m,n} + \delta^{ij} \delta_{m,n}.\;
\end{eqnarray}

%\begin{eqnarray}
%  \hat{\varphi}^i_n &=& \frac{1}{\sqrt{4\pi |n| g}}( \hat{a}^i_n +
%  \hat{a}^{i\dagger}_{-n}) \nonumber \\
%  \hat{\pi}^i_{n} &=& -i\sqrt{\pi |n| g}(\hat{a}^i_{-n} - \hat{a}^{i\dagger}_{n})\;,
%\end{eqnarray}

The Hamiltonian in terms of the deformed creation and annihilation operators is given by
\begin{equation}\label{HHAT}
  H = \frac{1}{2gR}\sum_i (\pi_0^i)^2 + \sum_{i,n\neq0}
  \frac{\omega_n}{2}(\hat{a}^{i\dagger}_n \hat{a}^i_n + \hat{a}^{i\dagger}_{-n} \hat{a}^i_{-n}),
\end{equation}
where $\omega_n=\frac{2\pi|n|}{R}$. Note that $\hat{a}^i_n$ and $\hat{a}^{j\dagger}_{n}$ are not standard oscillators, as one sees from (\ref{def-osc-1}) and (\ref{def-osc-2}), and hence nor is $\omega_n$
 the frequency of oscillation of the $n^{\textrm{th}}$ mode.

The generators of the modified $U(1)$ Kac-Moody algebra are
\begin{eqnarray}
  J_n^i = 
    \begin{cases}
      -i\sqrt{n} \; \hat{a}_n^i             & (n > 0) \\
      i\sqrt{-n} \; \hat{a}_{-n}^{i\dagger} & (n < 0)
    \end{cases},
&\hspace{1cm}& 
\Jb_n^i = 
    \begin{cases}
      -i\sqrt{n} \; \hat{a}_{-n}^i         & (n > 0) \\
      i\sqrt{-n} \; \hat{a}_{n}^{i\dagger} & (n < 0)
    \end{cases},
\end{eqnarray}
so that
\begin{eqnarray}
  \comm{J_m^i}{J_n^j} = \comm{\Jb_n^i}{\Jb_m^j} &=& \frac{i \pi g m^2}{R}   
  \theta^{ij} \delta_{m+n} + m\;\delta^{ij}\delta_{m+n} \label{JCR1}, \nonumber \\ && \nonumber \\
   \comm{J_m^i}{\Jb_n^j} &=& \frac{i \pi g m |m|}{R}  \theta^{ij} \delta_{m,n}. \label{JCR2}
\end{eqnarray}
Observe that new terms depending on $\theta$ have appeared in the commutation relations of the $U(1)$ Kac-Moody algebra.

Focusing on the non-zero mode terms of the Hamiltonian (\ref{HHAT}), for the time being, i.e., on 
\begin{equation}
\sum_{i,n\neq0}
  \frac{\omega_n}{2}(\hat{a}^{i\dagger}_n \hat{a}^i_n + \hat{a}^{i\dagger}_{-n} \hat{a}^i_{-n}),
\end{equation}
we may rewrite it using the new $J$'s and $\Jb$'s, so that
\begin{equation}
  \frac{2\pi}{R} \sum_{i,n>0} (J^i_{-n}\;J^i_n + \bar{J}^i_{-n}\;\bar{J}^i_n).
\end{equation}

The commutation relations (\ref{JCR2}) lead to the
relation
\begin{equation}\label{HJCR}
  \comm{H}{J^k_{-m}} = \frac{2\pi m}{R}\;J^k_{-m} -
  i\pi m^2\theta\epsilon_{ik}(J^i_{-m} + \bar{J}^i_m).
\end{equation}

Now we can write the ``conformal generators'' 
\begin{eqnarray}
  \hat{L}_0 &=& \frac{1}{2}\;J_0^i + \sum_{i,n>0} J^i_{-n}\;J^i_n \\
  \hat{L}_n &=& \frac{1}{2} \sum_{i,m} J^i_{n-m}\;J^i_m
  \hspace{5mm}(n\neq0) \\
  \hat{\bar{L}}_0 &=& \frac{1}{2}\;\hat{\bar{J}}_0^i + \sum_{i,n>0}
  \bar{J}^i_{-n}\;\bar{J}^i_n \\ 
  \hat{\bar{L}}_n &=& \frac{1}{2} \sum_{i,m} \bar{J}^i_{n-m}\;\bar{J}^i_m \hspace{5mm}(n\neq0)
\end{eqnarray}
where
\begin{equation}
  J^i_0 = \bar{J}^i_0 = \frac{\pi^i_0}{\sqrt{4\pi g}}.
\end{equation}

It should be noted that the generators $\hat{L}$ defined above do not close on a Virasoro-type algebra, i.e., $\comm{\hat{L}_m}{\hat{L}_n}$ is not proportional to $\hat{L}_{m+n}$ (plus a central term for $m+n=0$), because of the $\epsilon^{ij}$ term appearing in the RHS of equation (\ref{JCR2}).

The Hamiltonian (\ref{HHAT}) may finally be written as 
\begin{equation}
  H = \frac{2\pi}{R} (\hat{L}_0 + \hat{\bar{L}}_0).
\end{equation}

%%%%%%%%%%%%%%%%%%%%%%%%%%%%%%%%%%%%%%%%

\section{Deforming the $(3+1)$-d Theory}

%%%%%%%%%%%%%%%%%%%%%%%%%%%%%%%%%%%%%%%%

In this section we use the same procedure presented in the previous section for the case when we consider a different base space, say a $(3+1)$-d base space, while we still consider the same noncommutative target space as the case before, i.e., the noncommutative plane. This case will play a role in the next section, where we will use it to analyse how noncommutative target spaces may influence the black body radiation.

We start with a massless bosonic scalar field on the target space $\R^2$, but now with $(3+1)$-d base space, such that
\begin{equation}\label{eqNCTA:61}
\begin{array}{cccl}
\varphi:&S^1 \times S^1 \times S^1 \times \R &\longrightarrow& \R^2 \\
        &(\vec{x},t)         &\longmapsto    & \varphi(\vec{x},t).
\end{array}
\end{equation}
Observe that we are still considering compactified spatial coordinates, i.e., $S^1$, all of them with the same radius $R$.

The field components $\varphi^i(\vec{x},t)$ can be written in a Fourier series:
\begin{equation}
  \label{eqNCTA:62}
  \varphi^i(\vec{x},t) = \sum_{\vec{n}} \;e^{\frac{2\pi i}{R}\vec{n}\cdot \vec{x}} \; \varphi^i_{\vec{n}}(t)\;.
\end{equation}

Here we have $\vec{n}=(n_1,n_2,n_3)$, $n_i\in\mathbb{Z}$, and the Fourier components are written as
\begin{equation}
  \label{eqNCTA:63}
  \varphi^i_{\vec{n}}(t) = \frac{1}{R^3} \int d^3x \; e^{-\frac{2\pi i}{R}\vec{n}\cdot \vec{x}} \; \varphi^i(x,t)\;.
\end{equation}

The Lagrangian is given by
\begin{equation}
  \label{eqNCTA:64}
L = \frac{g}{2} \sum_i \int d^3x \big[ \big(\partial_t \varphi^i \big)^2 - \big(
  \nabla \varphi^i \big)^2 \big].
\end{equation}
The Lagrangian in terms of the Fourier modes of $\varphi^i$ is
\begin{equation}\label{eqNCTA:65}
  L =  \frac{gR^3}{2} \sum_{i} \bigg\{\dot{\varphi}^i_{\vec{n}} \dot{\varphi}^i_{-\vec{n}} -
  \bigg( \frac{2\pi |\vec{n}|}{R}\bigg)^2 \varphi^i_{\vec{n}} \varphi^i_{-\vec{n}} \bigg\}\;.
\end{equation}
The canonical momenta associated with the Fourier modes
$\dot{\varphi}^i_{\vec{n}}$ are
\begin{equation}
  \label{eqNCTA:66}
  \pi^i_{\vec{n}} = \frac{\partial L}{\partial \dot{\varphi}^i_{\vec{n}}} = gR^3\,\dot{\varphi}^i_{-\vec{n}}.
\end{equation}

Now we consider the noncommutative $\R^2$ as target space and follow the same procedure as the one we used in the previous section. Thus, the deformed fields are now
\begin{eqnarray}
\hat{\varphi}^{a}(\vec{x},t)&=&\varphi^{a}(\vec{x},t)-
\frac{1}{2}\epsilon^{ab}\theta \pi_{b}(\vec{x},t) \label{eqNCTA:67} \\
\hat{\pi}_{a}(\vec{x},t)&=&\pi_{a}(\vec{x},t)\;, \label{eqNCTA:68}
\end{eqnarray}
where the $\theta$ parameter has dimension of length square.

The commutation relation at equal time for the fields introduced above are
\begin{eqnarray}
\comm{\hat{\varphi}^{a}(\vec{x},t)}{\hat{\varphi}^{b}(\vec{y},t)}&=&i
\epsilon^{ab}\theta\delta(\vec{x}-\vec{y}), \label{eqNCTA:69}\\
\comm{\hat{\pi}_{a}(\vec{x},t)}{\hat{\pi}_{b}(\vec{y},t)}&=&0, \label{eqNCTA:69b}\\
\comm{\hat{\varphi}^{a}(\vec{x},t)}{\hat{\pi}_{b}(\vec{y},t)}&=&i\,\delta^a_b\delta(\vec{x}-\vec{y})\;.\label{eqNCTA:69c} 
\end{eqnarray}

We now replace the constant parameter $\theta$ by
\begin{equation}\label{eqNCTA:70}
\theta(\sigma)=\frac{\theta}{(\sqrt{2\pi}\sigma)^3} 
\;\exp \left[-
\sum_{i=1}^3 \frac{(x_i-y_i)^2}{2\sigma^2}
\right],
\end{equation}
where we make the simplifying assumptions $\sigma_1=\sigma_2=\sigma_3=\sigma$ in the more general expression $\sum_{i=1}^3\frac{(x_i-y_i)^2}{2\sigma_i^2}$ for the argument of the exponential, and $\theta(\sigma)$ has dimension $(\textrm{length})^{-1}$. The commutation relations (\ref{eqNCTA:69}) become now
\begin{equation}\label{eqNCTA:71}
\comm{\hat{\varphi}^{a}(\vec{x},t)}{\hat{\varphi}^{b}(\vec{y},t)}=\frac{i
\epsilon^{ab}\theta}{(\sqrt{2\pi}\sigma)^3}
\;\exp \left[- 
\sum^3_{i=1}\frac{(x_i-y_i)^2}{2\sigma^2}
\right]\;.\\ 
\end{equation}

The Fourier decomposition of the hatted fields $\hat{\varphi}^{a}$ are similar to those of the undeformed fields ${\varphi}^{a}$ in 
(\ref{eqNCTA:62}), i.e., 
\begin{equation} \label{eqNCTA:72}
\hat{\varphi}^{a}(\vec{x},t)=\sum_{\vec{n}}e^{\frac{2\pi i}{R}\vec{n}\cdot\vec{x}}\hat{\varphi}_{\vec{n}}^{a}.
\end{equation}
Using now
\begin{equation}
   \label{eqNCTA:73}
  \theta(\vec{n})=\theta\; e^{-\frac{2\pi^2\sigma^2 |\vec{n}|^2}{R^2}},
\end{equation}
we can write the deformation map for the Fourier modes of the fields as
\begin{eqnarray} 
\hat{\varphi}_{n}^{a}&=&\varphi_{n}^{a}-\frac{1}{2R^3}\epsilon^{ab}\theta(n)\pi_{-n}^{b}, \label{eqNCTA:74}\\
\hat{\pi}_n^{a}&=&\pi_n^{a}\;. \label{eqNCTA:75}
\end{eqnarray}

\begin{comment}
Observe that the dimension of $ \theta(\vec{n})$ is the same as that of $\theta(\sigma)$, i.e., length squared dimension.
\end{comment}

For the Fourier modes the commutation relations are
\begin{eqnarray}
 \comm{\hat{\varphi}_n^{a}}{\hat{\varphi}_m^{b}}&=&\frac{i\epsilon^{ab}\theta(n)}{R^3}\delta_{n+m,0}\\   \label{eqNCTA:76} 
\comm{\hat{\pi}_n^{a}}{\hat{\pi}_m^{b}}&=&0 \label{eqNCTA:77} \\
\comm{\hat{\varphi}_n^{a}}{\hat{\pi}_m^{b}}&=&i\,\delta^{ab} \delta_{mn} \label{eqNCTA:78}
\end{eqnarray}
and the Hamiltonian reads 
% \begin{eqnarray}
%   H &=& \frac{1}{2gR^3}\sum_{(i,\vec{n})} \; \hat{\pi}_{\vec{n}}^i\,\hat{\pi}_{-\vec{n}}^i +
%   (2\pi |\vec{n}| g R^2)^2
%   \, \hat{\varphi}^i_{\vec{n}} \, \hat{\varphi}^i_{-\vec{n}} \nonumber \\  
% % &=& \frac{1}{2gR}\sum_{(i,n)} \; \pi_n^i\,\pi_{-n}^i + (2\pi n g)^2
% %  (\varphi^i_n - \epsilon^{ij}\theta(n)R^{-1}\pi_{-n}^j)(\varphi^i_{-n} -
% %  \epsilon^{ij}\theta(n)R^{-1}\pi_{n}^j)  \nonumber \\
% &=& \frac{1}{2gR^3}  \sum_{(i,k,\vec{n})} \; \left[1+\left(\frac{\pi g |\vec{n}|
%       \theta(n)}{R}\right)^2\right]\,\pi_{\vec{n}}^i \, \pi_{-\vec{n}}^i + \nonumber \\
% & & \phantom{\frac{1}{2gR}  \sum_{(i,k,n)} \;} 
% + (2\pi |\vec{n}| g)^2 R^4\, \varphi_{\vec{n}}^i \, \varphi_{-\vec{n}}^i
% -(2\pi |\vec{n}| g)^2 R \epsilon_{ik}\theta(n)\varphi_{\vec{n}}^i \pi_{\vec{n}}^k \;.   \label{eq:HAMPHIPI3}
% \end{eqnarray}

% \begin{equation}
%   \Omega^2(\vec{n}) = 1+\left(\frac{\pi g |\vec{n}|\theta(n)}{R})\right)^2\hspace{10mm}\mathrm{and}\hspace{10mm}\omega_{\vec{n}} = \frac{2
%     \pi |\vec{n}|}{R}\;.
% \end{equation}
\begin{eqnarray}\label{eqNCTA:79}
  H &=& H_0+\sum_{i,\vec{n}} \left(\frac{\Omega^2_{\vec{n}}}{2gR^3} \;
  \pi_{\vec{n}}^i\,\pi_{-\vec{n}}^i + 
  \frac{g R^3}{2}\;\omega_{\vec{n}}^2 \,\varphi^i_{\vec{n}} \,
  \varphi^i_{-\vec{n}} -
  \frac{g}{2}\;\omega^2_{\vec{n}}\;\theta(n)\;\epsilon_{ik}\,\varphi^i_{\vec{n}}\,\pi^k_{\vec{n}}\right) \nonumber \\ && 
\end{eqnarray}
where 
\begin{equation}\label{eqNCTA:80}
   \Omega^2_{\vec{n}} = 1+\left(\frac{\pi g |\vec{n}|\theta(n)}{R}\right)^2
\end{equation}
and
\begin{equation}\label{eqNCTA:81}
\omega_{\vec{n}} = \frac{2 \pi |\vec{n}|}{R}\;.
\end{equation}

We can again apply the Schwinger procedure. We use the operators $A^i_{\vec{n}}$'s  and
$A^{i\;\dagger}_{\vec{n}}$'s with $i=1,2$, defined similarly to (\ref{eq:117}), so that the Hamiltonian may be written as
 \begin{eqnarray}
  H &=& 
%\frac{(\Pi_0)^2 + (\tilde{\Pi}_0)^2}{4gR} + 
H_0+ \sum_{\vec{n} \neq 0} \omega_{\vec{n}} 
\bigg\{ 
\left( \Omega_{\vec{n}}  + \frac{\pi|\vec{n}| g \theta(n)}{R} \right)
{A_{\vec{n}}^1}^\dagger A_{\vec{n}}^1 + 
\left( \Omega_{\vec{n}}  - \frac{\pi|\vec{n}| g \theta(n)}{R} \right)
{A_{\vec{n}}^2}^\dagger A_{\vec{n}}^2 \bigg\} \nonumber \\ && \nonumber \\
 &=& 
%\frac{(\Pi_0)^2 + (\tilde{\Pi}_0)^2}{4gR} + 
\sum_{\vec{n} \neq 0} \omega_{\vec{n}} 
\bigg\{ 
\Lambda^1_{\vec{n}} \, {A_{\vec{n}}^1}^\dagger A_{\vec{n}}^1 +
\Lambda^2_{\vec{n}} \, {A_{\vec{n}}^2}^\dagger A_{\vec{n}}^2
\bigg\}\label{eqNCTA:82}
\end{eqnarray}
with
\begin{eqnarray}
  \Lambda_{\vec{n}}^1 &=&  \Omega_{\vec{n}}  + \frac{\pi|\vec{n}| g \theta(n)}{R} \label{eqNCTA:83}\\
  \Lambda_{\vec{n}}^2 &=&  \Omega_{\vec{n}}  - \frac{\pi|\vec{n}| g \theta(n)}{R}\;. \label{eqNCTA:84}
\end{eqnarray}

We observe that the splitting of the energy levels of the oscillators are still present in this case.

%%%%%%%%%%%%%%%%%%%%%%%%%%%%%%

\section{Deformed Black Body Radiation}

%%%%%%%%%%%%%%%%%%%%%%%%%%%%%%

In this section we analyse the black body radiation related with the quantum field theory presented in the last section. We have shown that there is a splitting of the energy levels, equations (\ref{eqNCTA:83}) and (\ref{eqNCTA:84}), which necessarily implies a modification of the dispersion relations of the theory. Here we show how these deformed dispersion relations affect the black body radiation. 

Using the Hamiltonian (\ref{eqNCTA:82}), we can write the partition function as\footnote{We ignore the zero mode. It is not relevant, since it is associated with the overall translation of the system.}
\begin{eqnarray}
  \label{eq:Z3}
  Z &=& \tr e^{-\beta \; H} = \prod_{\vec{k}\neq0} \sum_{m,n=0}^{\infty} \;
  e^{-\beta \; \omega_{\vec{k}} \; \Lambda_{\vec{k}}^1 \; n} \; 
  e^{-\beta \; \omega_{\vec{k}}  \; \Lambda_{\vec{k}}^2 \; m} \nonumber \\
&=& \prod_{\vec{k} \neq 0} \left( \frac{1}{1-e^{-\beta \omega_{\vec{k}} \Lambda_{\vec{k}}^1}} \right)
\left(  \frac{1}{1-e^{-\beta \omega_{\vec{k}} \Lambda_{\vec{k}}^2}}   \right).
\end{eqnarray}
(We ignore the zero mode. It is not relevant, being associated with the overall translation of the system.) From this partition function we may consider the energy in a finite volume $V\in\R^3$ of the system defined by
\begin{equation}
  \label{eq:U}
  U = - \frac{\partial}{\partial \beta} \; \ln Z.
\end{equation}
Noting that
\begin{equation}
  \label{eq:LNZ}
  \ln Z = - \sum_{\vec{k}\neq 0} \left[ \ln \left( 1 - e^{-\beta
        \omega_{\vec{k}} \Lambda_{\vec{k}}^1} \right) + \ln \left( 1 - e^{-\beta \omega_{\vec{k}} \Lambda_{\vec{k}}^2}\right) \right],
\end{equation}
we obtain
\begin{eqnarray}
  U &=& \sum_{\vec{k}\neq 0} \left[
    \frac{\omega_{\vec{k}}\;\Lambda_{\vec{k}}^1}{(e^{\beta\;\omega_{\vec{k}}\;\Lambda_{\vec{k}}^1} - 1)}  + \frac{\omega_{\vec{k}}\;\Lambda_{\vec{k}}^2}{(e^{\beta\;\omega_{\vec{k}}\;\Lambda_{\vec{k}}^2} - 1)} 
\right] \nonumber \\
&=& 2 \; \sum_{\vec{k} =1}^{\infty} \left[
    \frac{\omega_{\vec{k}}\;\Lambda_{\vec{k}}^1}{(e^{\beta\;\omega_{\vec{k}}\;\Lambda_{\vec{k}}^1} - 1)}  + \frac{\omega_{\vec{k}}\;\Lambda_{\vec{k}}^2}{(e^{\beta\;\omega_{\vec{k}}\;\Lambda_{\vec{k}}^2} - 1)} \right] \label{eq:UT}\\
&=& U_1 + U_2. \nonumber 
\end{eqnarray}
Each of the terms $U_1$ and $U_2$ corresponds to the energy of the oscillators.

The limit to infinite volume for the system is obtained by considering
\begin{eqnarray}
  \label{eq:Ui}
  U_i &=& \frac{2R^3}{(2\pi)^3} \int_0^\infty dk \; (4\pi k^2)\frac{k
    \Lambda^i(k)}{e^{\beta\;k\;\Lambda^i(k)}-1} \nonumber \\
&=& \frac{V}{\pi^2} \int_0^\infty dk \frac{k^3 \Lambda^i(k)}{e^{\beta\;k\;\Lambda^i(k)}-1},
\end{eqnarray}
with $i=1,2$, where now\footnote{We adopt the choice $g=1/4\pi$.}
\begin{eqnarray}
  \Lambda^1(k) &=&  \sqrt{1 +\left(\frac{k\,\theta(k)}{8\pi} \right)^2} + \frac{k\,\theta(k)}{8\pi}, \\
  \Lambda^2(k) &=&  \sqrt{1 +\left(\frac{k\,\theta(k)}{8\pi} \right)^2} - \frac{k\,\theta(k)}{8\pi}\;,
\end{eqnarray}
where
\begin{equation}
  \theta(k)=\theta \; e^{-\frac{1}{2} \,\sigma^2 \, k^2}\;.
\end{equation}
In units $c=\hbar=1$, we have $k=\omega=\textrm{frequency}$ (this $k$ is the wave number,
not an integer), so that we can automatically rewrite the energy density as an integral
in $\omega$, i.e., 
\begin{eqnarray}
  \label{eq:UV}
  \frac{U}{V} &=& \frac{1}{\pi^2}  \int_0^\infty d\omega \left[ \frac{\omega^3\;
      \Lambda^1(k)}{e^{\beta\;\omega\;\Lambda^1(k)}-1} + \frac{\omega^3 \Lambda^2(k)}{e^{\beta\;\omega\;\Lambda^2(k)}-1} \right]
\end{eqnarray}

In the limit $\theta \rightarrow 0$, $\Lambda^i(k) \rightarrow 1$ and the
energy density reduces to
\begin{eqnarray}
  \label{eq:UV0}
  \frac{U}{V} = \frac{2}{\pi^2}  \int_0^\infty d\omega \;
  \frac{\omega^3}{e^{\beta\;\omega}-1} = \int_0^\infty d\omega \; u(\omega),
\end{eqnarray}
with $u(\omega)$ being the Planck distribution, which is such that after integration it
gives $U/V \propto T^4$.

We may calculate the expansion of equation (\ref{eq:UT}) in powers of
$\theta$, using the usual normalization:
\begin{eqnarray}
  \label{eq:UVEXP}
  U(\theta) &=& U(0) + \frac{dU}{d\theta}\bigg|_{\theta=0} \theta +
  \frac{d^2U}{d\theta^2}\bigg|_{\theta=0} \theta^2 + \cdots
\end{eqnarray}

The first term $U(0)$ of the expansion is given by equation
(\ref{eq:UV0}). The term of order $\theta$ is zero and the term of order
$\theta^2$ is
\begin{equation}
  \label{eq:EXP2}
  \frac{d^2U}{d\theta^2}\bigg|_{\theta=0} \theta^2 =  \frac{1}{\pi^2}
  \int_0^\infty d\omega \; \frac{\omega^5 \;
    \theta^2\;F(\sigma,\beta,\omega)}{64(e^{\beta \omega} - 1)^3 \pi^2}
\end{equation}
where
\begin{eqnarray}
  F(\sigma,\beta,\omega) &=& e^{\omega(-\sigma^2\omega+2\beta)} -2e^{\omega(-\sigma^2\omega+\beta)}
    + e^{-\sigma^2\omega^2} - 3 \omega  e^{\omega(-\sigma^2
      \omega+2\beta)} \beta + \nonumber \\ && \nonumber \\
&& + 3 \omega e^{\omega (-\sigma^2 \omega+\beta)}\beta +\omega^2
e^{\omega(-\sigma^2 \omega+2\beta)} \beta^2  
+ \omega^2 e^{\omega (-\sigma^2 \omega+\beta)}\beta^2   
\end{eqnarray}

The correction, due to noncommutativity, for the black body
radiation is not a polynomial in $\beta$. 

So that we may obtain some intuition to help us to better understand the behaviour of the corrections, we present
some graphs in what follows. These are graphs that are to be compared with the well-known black body radiation graph. 

It is remarkable that the behaviour of the radiation obtained after
introducing noncommutativity is similar to that of a
regularisation by the parameters $\theta$ and $\sigma$. For
phenomenological reasons $\theta$ cannot
be too big, nor can $\sigma$ be too small as that would make frequencies increase too fast with $|\vec{n}|$ (see below).

\begin{figure}[h!]
\begin{center}
\includegraphics[scale=0.4]{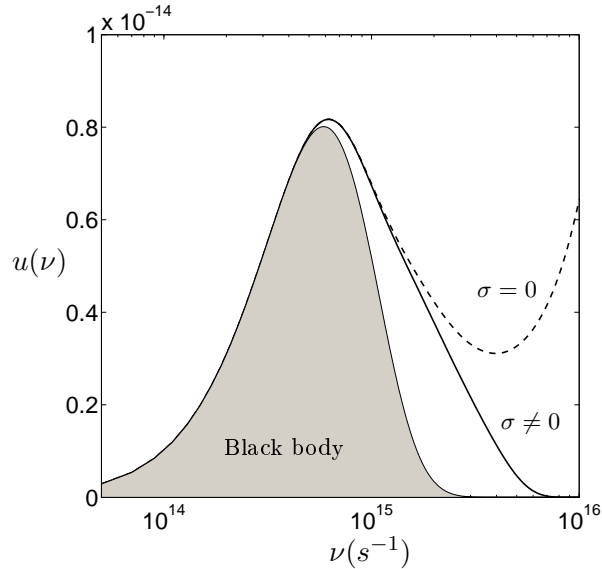}
\begin{picture}(0,0)
\put(-160,40){\footnotesize Black body}
\put(-64,100){\footnotesize $\sigma=0$}
\put(-55,50){\footnotesize $\sigma \neq 0$}
\put(-120,0){$\nu(s^{-1})$}
\put(-240,110){$u(\nu)$}
\end{picture}
\caption{\small The shady region corresponds to the black body radiation with temperature $T=1\times 10^4$\;K. The solid line delimits the region of the new radiation, which is given by the deformed theory for $\sigma=1\times
  10^{-15}$. The dashed line gives the radiation curves of the deformed theory with null $\sigma$. Observe that in this case, there is a divergence in the energy density at high frequencies.}
\label{FIG1}
\end{center}
\end{figure}

In the graph of figure \ref{FIG1}, we see the radiation curves for a temperature of $1\times 10^4$\;K, both for the usual black body distribution and for its deformed distribution. In the case of the deformed theory, but with $\sigma=0$, we clearly see a divergence in the energy density when the frequency of the radiation is increased. This is because when $\sigma=0$, the oscillator frequencies $\omega_{\vec{n}}\left( \Omega_{\vec{n}}\mp\frac{\pi|\vec{n}|g\theta(n)}{R}\right)$ increase too fast, that is, like $|\vec{n}|^3$ as $|\vec{n}|\to\infty$. For the (arbitrary) choice $\sigma=1\times
  10^{-15}$m, in the figures, we see a deviation of the radiation energy density curve for the deformed theory from the usual black body radiation.

\begin{figure}[h!]
\begin{center}
\includegraphics[scale=0.32]{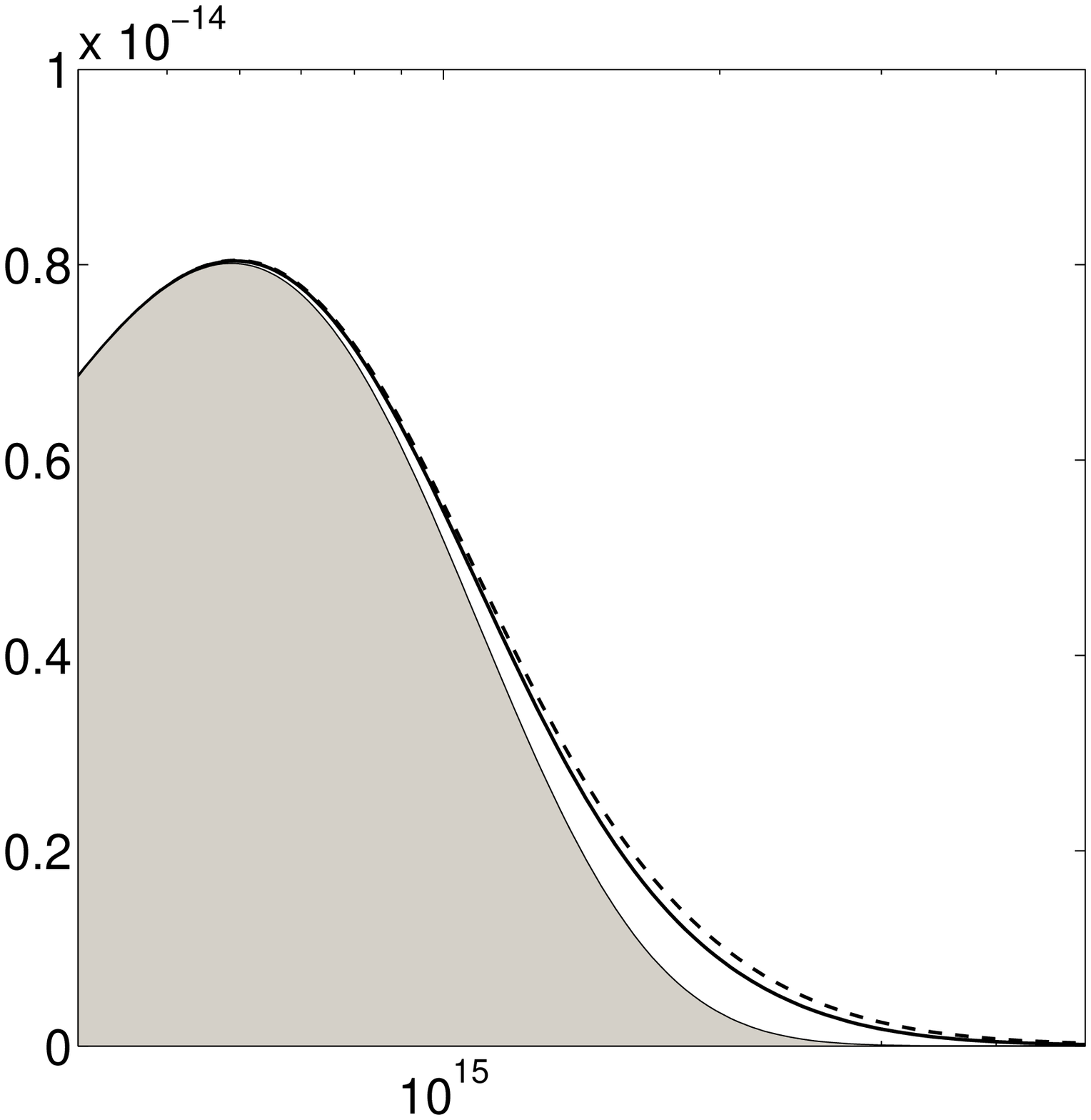}
\includegraphics[scale=0.32]{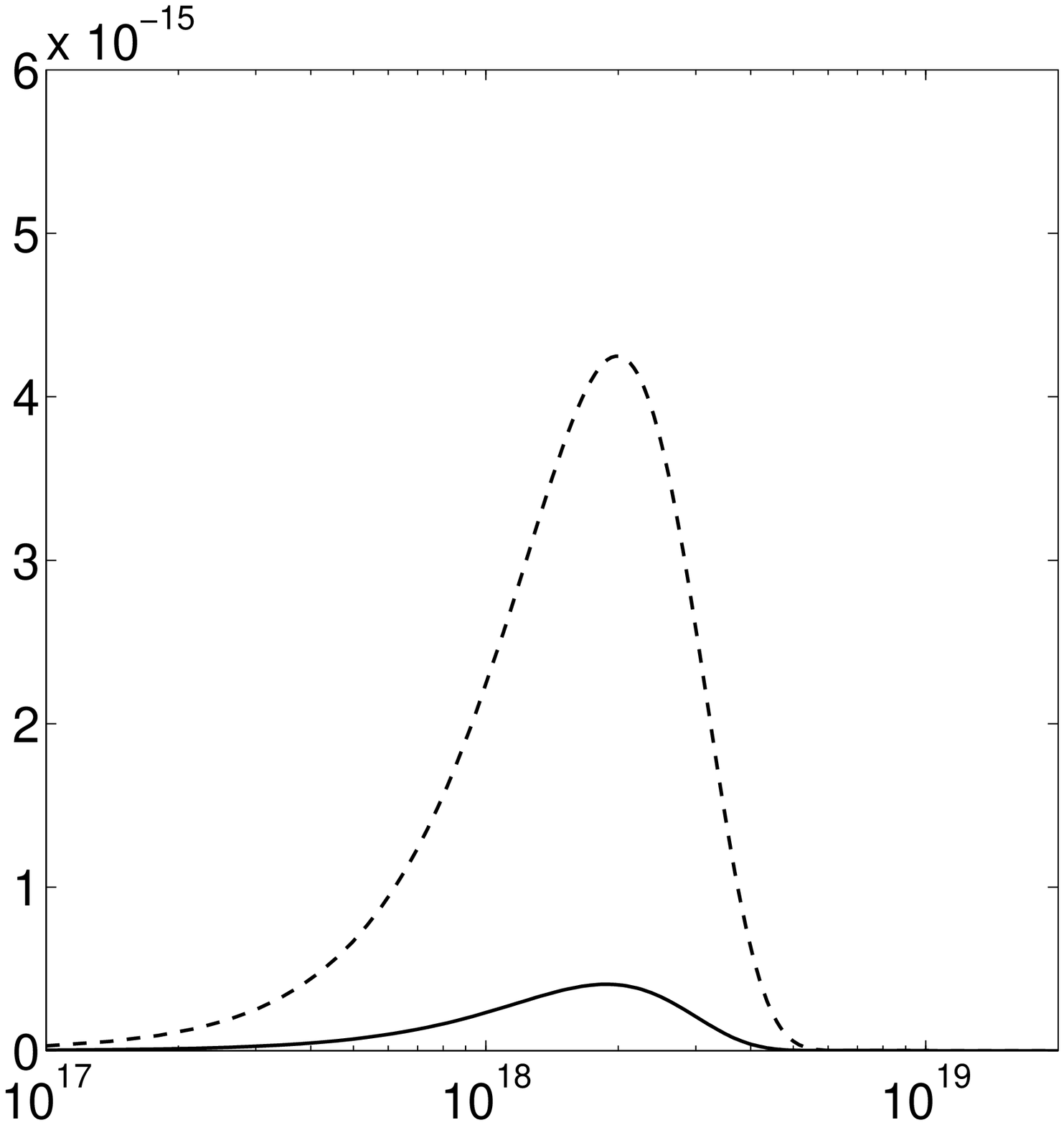}
\begin{picture}(0,0)
\put(-345,40){\footnotesize Black body}
\put(-95,122){\scriptsize $\theta=1.9 \times 10^{-14}$m}
\put(-95,33){\scriptsize $\theta=1.7 \times 10^{-14}$m}
\put(-290,170){\small \textbf{(a)}}
\put(-95,170){\small \textbf{(b)}}
\put(-290,0){$\nu(s^{-1})$}
\put(-390,80){$u(\nu)$}
\put(-95,0){$\nu(s^{-1})$}
\put(-198,80){$u(\nu)$}
\end{picture}
\caption{\small \textbf{(a)} When $\sigma$ is nonnull ($\sigma = 1 \times
  10^{-15}$ and $T=1\times 10^4$K in both graphs), there exists a very small deviation with respect to the usual black body radiation at low frequencies. The graph shows the energy density curve for the new radiation with $\theta = 1.7 \times 10^{-14}$
  (solid line) and $\theta =
  1.9 \times 10^{-14}$ (dashed line); \;\;\textbf{(b)} A not-so-small deviation is found for high frequencies.}  
\label{FIG2}
\end{center}
\end{figure}

The effects described above are inferred from low frequencies. The effective deviation with respect to the black body radiation is regulated by $\sigma$ and $\theta$. One can see the behaviour of such deviations either in low frequencies or in high frequencies in Figure \ref{FIG2}. In these figures the temperature is $1\times
10^4$\;K with fixed $\sigma = 1 \times 10^{-18}$m.

One can notice that for low frequencies, the effects can made very small by choosing values for the $\theta$ parameter. The solid and dashed curves correspond to the values $\theta=1.7 \times
10^{-14}$ and $\theta=1.9 \times 10^{-14}$ respectively. 

On the other hand, for high frequencies one can notice that a modification appears in the radiation. Such a modification is given by a peak in the radiation curve, which we refer to as the secondary peak. We observe in Figure \ref{FIG2} how distinct values of $\theta$ lead to different values for the highest amplitude of the secondary peak.

\begin{figure}[t!]
\begin{center}
\includegraphics[scale=0.6]{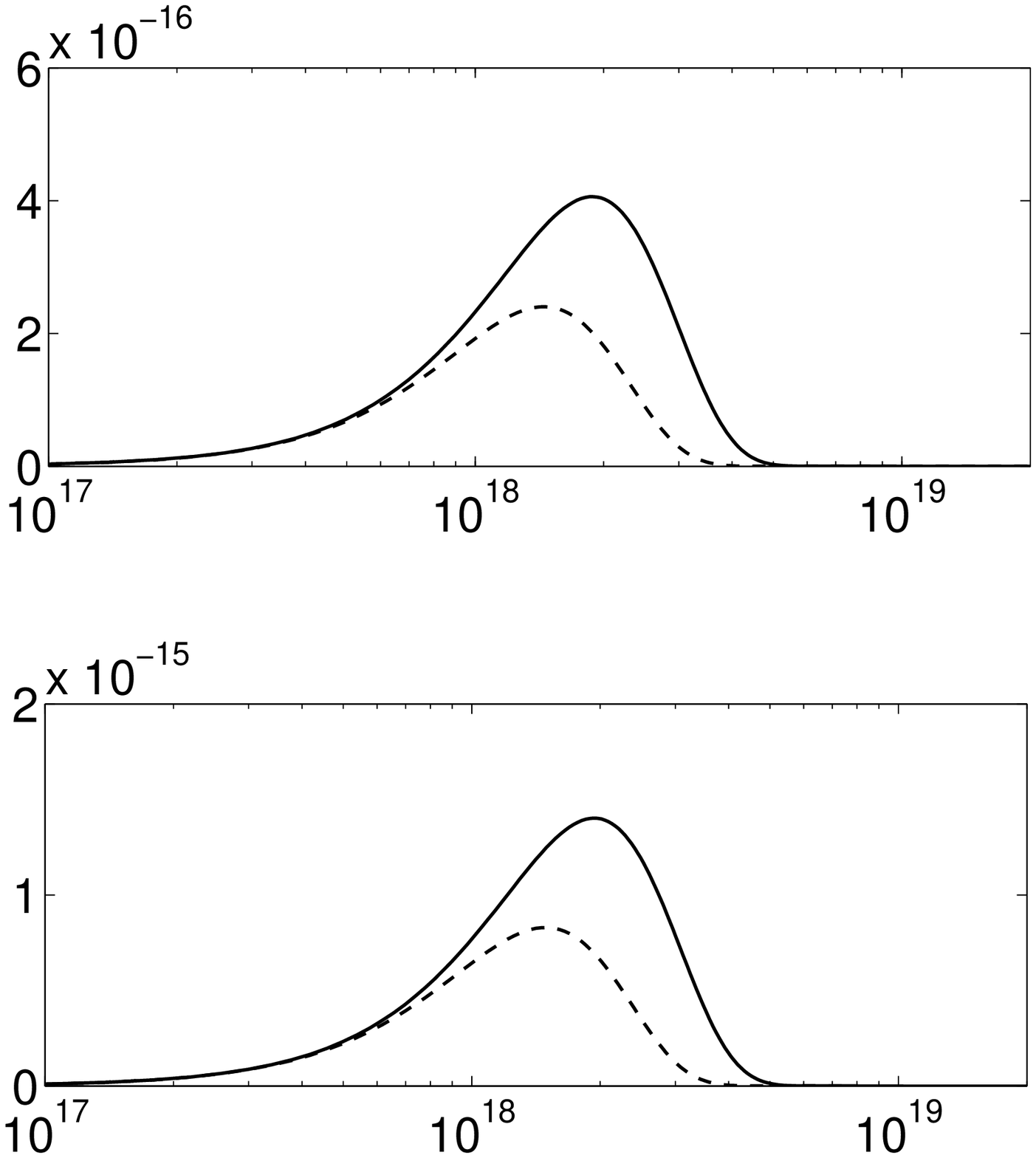}
\begin{picture}(0,0)
\put(-280,290){\scriptsize $\theta=1.7 \times 10^{-14}$m}
\put(-280,110){\scriptsize $\theta=1.8 \times 10^{-14}$m}
\put(-140,90){\scriptsize $\sigma=1.0 \times 10^{-18}$m}
\put(-148,55){\scriptsize $\sigma=1.3 \times 10^{-18}$m}
\put(-143,265){\scriptsize $\sigma=1.0 \times 10^{-18}$m}
\put(-151,230){\scriptsize $\sigma=1.3 \times 10^{-18}$m}
\put(-185,-5){$\nu(s^{-1})$}
\put(-345,70){$u(\nu)$}
\put(-185,165){$\nu (s^{-1})$}
\put(-345,245){$u(\nu)$}
\end{picture}
\caption{\small Another example that the frequency and the highest amplitude depend on $\sigma$ and  $\theta$. The dependence of the highest amplitude on $\sigma$ is stronger than the dependence on $\theta$. For both graphs, $T=1\times 10^4$K.}
\label{FIG3}
\end{center}
\end{figure}

The frequency and the highest amplitude of the secondary peak are functions of the parameters. This can be seen in Figure \ref{FIG3}. The growth of $\sigma$ leads to the decrease of the highest amplitude of the modification, as well as the translation of the secondary peak to low frequencies. This agrees with the fact that when $\sigma$ goes to zero, the highest amplitude grows, and the divergence appears.

One can note also in Figure \ref{FIG3} that as $\theta$ grows, the highest amplitude of the modification grows as well. The value of $\theta$ is also given by the position of the peak of the modification, however such dependence is stronger on $\sigma$ than on $\theta$.

\section{Conclusion}

This work is based on the deformation of the target space of a field in a quantum field theory by considering the noncommutative plane $\R^2$ as target space. It is based on the earlier work of \cite{Carmona:2003kh,Carmona:2002iv}. One of the simplest of the quantum field theories, i.e., the free massless boson field theory was considered in two different cases. In the first case we recalled this theory when the base space is a cylinder $S^1\times \R$. In the second case the base space is $(S^1)^3\times\R$.

A first consequence of such a deformation is the appearance of a term proportional to a component of angular momentum in the Hamiltonian of the theory. It affects the split of the energy levels. This split deforms the dispersion relations of the theory, which have consequences for the radiation spectrum of this theory. A new feature in our treatment is that besides the noncommutativity parameter $\theta$, we also introduce another parameter, $\sigma$, in the model, resulting in the commutation relation (\ref{eq:NCCRSIGMA}). The radiation
spectrum depends on both these parameters. In order to understand its
dependence on these parameters, we expanded the energy density upto second
order in theta and graphically studied its dependence on the
parameters. The analysis of these graphs reveals that the deviation of the new radiation spectrum with respect to the black body radiation is stronger in $\sigma$ than in $\theta$.

As an application of the deformed black body spectrum, one may calculate the GZK cut-off. In \cite{Geddes:1995sd}, the GZK cut-off is calculated using the distribution of black body radiation. A follow up of the present work is to evaluate a new GZK cut-off using the deformed black body distribution obtained in this work.

\section*{Acknowledgments}

The authors thank Prof. R. Abramo for pointing out that the deformed black body radiation may be relevant for the GZK cut-off. The authors also thank Prof. Y. Albuquerque for pointing out the work \cite{Geddes:1995sd}. We also thank Prof. J. Gamboa and Prof. J. Cortez for calling our attention to \cite{Carmona:2002iv,Carmona:2003kh,Das:2004vc,Gamboa:2005pd,Falomir:2005it,Gamboa:2005bf,Arias:2006fu,Arias:2007xt}, which already contain
many of our ideas. We have tried to acknowledge their contributions
properly in the text. The work of A.P.B. is supported in part by US Department of Energy under grant number DE-FG02-85ER40231. 

\bibliographystyle{JHEP}
\bibliography{bib_abelian}

\end{document}